\documentclass[twocolumn]{aastex63}
\usepackage{natbib}
\usepackage{needspace}
\usepackage{tabularx}
\usepackage{booktabs}
\usepackage{multirow}
\usepackage{amsmath}
\usepackage{amstext}
\usepackage{mathptmx}
\usepackage{tabularx} 
\usepackage{hyperref}
\usepackage{tablefootnote}
\usepackage{placeins}
\let\tablenum\relax % Let siunitx redefine tablenum
\usepackage{siunitx}

\usepackage{amsmath}
\usepackage{mathptmx}
\usepackage{txfonts}
\usepackage{graphicx}
\usepackage{placeins}
\usepackage{multirow}
\usepackage{array}
\usepackage{natbib}
\usepackage{tabularx}
\usepackage{comment}

\usepackage[T1]{fontenc}
\usepackage{ae,aecompl}
\usepackage{lineno}
\usepackage{listings}

\usepackage{layout}

\usepackage{graphicx}
\newcommand*\patchAmsMathEnvironmentForLineno[1]{
  \expandafter\let\csname old#1\expandafter\endcsname\csname #1\endcsname
  \expandafter\let\csname oldend#1\expandafter\endcsname\csname end#1\endcsname
  \renewenvironment{#1}
  {\linenomath\csname old#1\endcsname}
  {\csname oldend#1\endcsname\endlinenomath}}    \newcommand*\patchBothAmsMathEnvironmentsForLineno[1]{
  \patchAmsMathEnvironmentForLineno{#1}
  \patchAmsMathEnvironmentForLineno{#1*}}
  \AtBeginDocument{
  \patchBothAmsMathEnvironmentsForLineno{equation}
  \patchBothAmsMathEnvironmentsForLineno{align}
  \patchBothAmsMathEnvironmentsForLineno{flalign}
  \patchBothAmsMathEnvironmentsForLineno{alignat}
  \patchBothAmsMathEnvironmentsForLineno{gather}
  \patchBothAmsMathEnvironmentsForLineno{multline}
}

\newcommand{\target}{HD~189733}
\newcommand{\planet}{HD~189733b}

\shortauthors{Ehrich, Dittmann et al.}
\begin{document}
\title{An H$\alpha$ Transit of HD 189733b to Assess Stellar Activity Across the Transit Chord Close to JWST Observations}

\author[0009-0002-6199-9041]{Kingsley E.~Ehrich}
\email{kehrich@ufl.edu}
\affiliation{Department of Astronomy, University of Florida, Bryant Space Science Center, Stadium Road, Gainesville, FL 32611, USA }

\author[0000-0001-7730-2240]{Jason A. Dittmann}
\affiliation{Department of Astronomy, University of Florida, Bryant Space Science Center, Stadium Road, Gainesville, FL 32611, USA }

\author[0000-0003-1312-9391]{Samuel P. Halverson}
\affiliation{Jet Propulsion Laboratory, California Institute of Technology, 4800 Oak Grove Drive, Pasadena, CA 91109, USA}

\author[0000-0002-3516-3446]{Alejandro Camaz\'on-Pinilla}
\affiliation{Department of Astronomy, University of Florida, Bryant Space Science Center, Stadium Road, Gainesville, FL 32611, USA }

\begin{abstract}

Transmission spectroscopy allows us to detect molecules in planetary atmospheres, but is subject to contamination from inhomogeneities on the stellar surface. Quantifying the extent of this contamination is essential for accurate measurements of atmospheric composition, as stellar activity can manifest as false atmospheric signals in planetary transmission spectra. We present a study of hot Jupiter {\planet}, which has over 50 hours of JWST observations scheduled or taken, to measure the activity level of the host star at the current epoch. We utilize high-resolution spectra of the H$\alpha$ line from the MEGARA spectrograph on the 10-m GTC to examine the activity level of {\target} during a transit. We measure H$\alpha$ becoming shallower mid-transit by an H$\alpha$ index of $\delta = 0.00156 \pm 0.00026$, which suggests that {\planet} crosses an active region as it transits. We posit this deviation is likely caused by a spot along the transit chord with an approximate radius of $ R_\mathrm{spot} = 3.47 \pm 0.30 R_{\oplus}$ becoming occulted during transit. Including an approximation for unocculted spots, we estimate that this spot could result in transit depth variations of $\sim$17 ppm at the 4.3 micron CO$_2$ feature. Since this is comparable to JWST NIRCam Grism mode’s noise floor of $\sim$20 ppm, it could bias atmospheric studies by altering the inferred depths of the planet’s features. Thus, we suggest ground-based high-resolution monitoring of activity indicator species concurrently taken with JWST data when feasible to disentangle stellar activity signals from planetary atmospheric signals during transit.

\end{abstract}


\section{Introduction}
\label{sec:intro}

Transmission spectroscopy allows for characterization of planetary and atmospheric composition. However, challenges associated with ground-based observations, such as variable telluric contamination, changing observing conditions, and thermal background, can place limits on the systematics floor for exoplanet atmospheric measurements. Because of this, space-based spectroscopic studies of exoplanets are appealing. The tradeoff with space-based spectroscopy is that although it is impervious to several of these issues, resolution and/or collecting area is sacrificed due to practical constraints of space-based observatories. An issue with spectroscopy, however, is that the planet being studied is typically unresolved. As a result, stellar signals from the host star become inevitably entangled with the planetary signals, systematically contaminating the exoplanet's transmission spectrum \citep{2008MNRAS.385..109P}. There is a wavelength-dependent source of contamination called the Transit Light Source Effect. This arises because the assumed light source during transit is the pre-transit observable stellar disk. However, stellar inhomogeneities along the surface create a different spectral signature due to different contributions of spots and faculae to the transmission spectra \citep{Rackham2018Feb}. Thus, the disk-averaged spectrum of the star mid-transit is different than it was pre-transit, since the planet covers different parts of the star along the transit chord, preferentially weighting different areas of the inhomogeneous stellar surface. 

The composition of star spots themselves can also present its own issue. Since spots are cooler regions of the star than the disk-averaged photosphere, molecules can form and exist in them (\citealp{Wohl1971Feb}; \citealp{1995ApJ...452..879N}; \citealp{2003A&A...412..513B}). If there is a molecule present in the planet's atmosphere that is also present in a spot on its host star, unocculted spots can deepen absorption features in the planet's transmission spectrum \citep{Thompson_2024}. In some cases, the stellar signal from inhomogeneities on the stellar surface can overpower the signal from the planet, and temporal variation on the stellar surface can cause stellar lines to evolve over time \citep{Rackham2023Jan}. Where heavy stellar contamination is present, neglecting to account for stellar activity can lead to an underestimation of H$_2$O abundance by over 2 orders of magnitude \citep{Thompson_2024}. Unocculted spots can be approximated  as static during a transit, since the transit timescale is on the order of hours, while the spot evolution timescale can be much longer than that (weeks to months) \citep{Giles2017Dec}. With this in mind, a general correction for the wavelength-dependent contribution to the transmission spectrum may be sufficient \citep{2011MNRAS.416.1443S}. These unocculted spots cause an overall dimming effect. Unocculted spots can also introduce a slope to the inferred transmission spectrum, even in the absence of planetary haze, further complicating atmospheric retrievals and analysis of mixed stellar and planetary transmission spectra. {\planet} has a strongly sloped transmission spectrum, which could be due to unocculted spots alone if their covering fraction were only $\sim$5.6 percent \citep{2014ApJ...791...55M}. The inferred unocculted spot covering fraction is estimated to be $\sim$1-1.7 percent based on rotational variability monitoring (\citealp{2008MNRAS.385..109P}; \citealp{2013MNRAS.432.2917P}; \citealp{2009A&A...505..891S}; \citealp{2011MNRAS.416.1443S}). Furthermore, underestimating spot contribution can lead to inaccurate measurements of hazes within planetary atmospheres. Thus, corrections for spot contributions based on stellar variability should be treated as a lower limit, and may lead to underestimated corrections in transmission spectra.

Occulted spots, however, typically have opposite effects compared to unocculted spots. \citealt{Komori_2025} and \citealt{2005asus.book.....W} show that in general, unocculted spots can either cause absorption lines to become deeper or shallower than they are in the photosphere. Therefore, both outcomes are also possible in case of occulted spots. Section 4.2.5 of \citealp{2025AJ....170...61M} shows that occulted spots can have varying effects on transmission spectrum features, as they can cause them to become both deeper and shallower depending on the species, temperatures, and latitudes involved. \citealt{article} note that while larger spot-crossing events are relatively easily noticeable, smaller spot-crossing events may be nearly undetectable in broad photometric bandpasses but can have a non-negligible effect on molecular detections. These effects are exacerbated when multiple transit visits are averaged together to increase signal. \citealt{article} suggest that high-resolution spectroscopy and cross-correlation techniques are potential methods to diagnose what kind of stellar activity is present, and the nature of its spectra.
Furthermore, \citealt{article} performed analysis to examine the effects of unocculted spots on spectra. They compared spectra of a modeled spotted M-dwarf surface to that of the same M-dwarf with the spots rotated off the stellar surface by taking a ratio of the spectrum from the spotted surface divided by the spectrum of the non-spotted surface. They found that some molecular species have stronger absorption in the spot versus quiet photosphere, whereas other lines present the opposite result.  It is not obvious which effect will dominate for a crossing event for bright or dark active regions viewed in the H$\alpha$ line center, so we perform an empirical test using solar H$\alpha$ data in Section \ref{sec:discussion}.

Both occulted and unocculted spots are relevant to this work. While we measure the activity level along the transit chord, the scale of activity present there may be indicative of the relative activity level of the overall stellar disk. Both unocculted and occulted spots, as well as other kinds of stellar activity such as brighter plage regions also associated with spots and magnetic activity, may interfere with transmission spectroscopy measurements and result in biases in the inferences of molecular abundances present in the planet's atmosphere.

The paper is organized into the following sections: Section 2, which provides context to our selected target; Section 3, which details our observations and data reduction process; Section 4, which covers our analysis procedures; Section 5, where we discuss the results; and Section 6, where we summarize our primary findings.

% ------- Break into separate section?
\section{The Stellar Activity History of HD 189733b}
HD 189733 is a bright (V = 7.7) star known to host a transiting hot Jupiter planet \citep{Bouchy2005Dec}.  The planet has a orbital period of 2.2 days, and a radius of 1.13 $R_J$ \citep{Bouchy2005Dec}. {\target} is a fairly active K2V star (\citealp{Bouchy2005Dec}, \citealp{Wright2004Jun}). There is evidence for stellar contamination in prior transmission spectra taken of {\planet}; for example, the velocity structure of the H$\alpha$ line in the planet's transmission spectrum in previous transit data indicates that its excess absorption originates from the host star \citep{2016MNRAS.462.1012B}. \citealp{Cauley2017Apr} find that the strongest absorption of H$\alpha$ occurs when the star is most active. \citealp{2016MNRAS.462.1012B} examined three separate transits with concurrent observations of H$\alpha$ and Ca II H \& K. The H$\alpha$ absorption traced out the same signal as the Ca II H \& K line, suggesting that the variations in H$\alpha$ were indeed stellar in origin. They detected excess absorption in the H$\alpha$ transit depth of $\delta = 0.0074 \pm 0.0044$ and $\delta = 0.0214 \pm 0.0022$ for the first and second transits, respectively. A flare occurred during their third transit which prevented an accurate measure of transit depth due to emission occurring in the line cores of both activity indicators. With this in mind, it is clear that stellar contamination has posed a risk for this target in the past and will likely continue to do so in the future. 

The system's brightness makes it a prime candidate for exosphere characterization, thus, it is the subject of several James Webb Space Telescope (JWST) programs totaling nearly 58 hours in Cycle 1 GO. JWST's space-based spectroscopy has a uniquely wide window into planetary atmospheres since its infrared wavelength range is not limited by typical challenges ground-based instruments face, such as telluric contamination. However, spectra from JWST's MIRI are limited to low- and medium-resolution studies, since it has a maximum spectral resolving power of $\sim$3700 \citep{2023A&A...675A.111A}. Supplemental high-resolution data taken concurrently with JWST data has the potential to be highly beneficial for this reason. Particularly, nuances of stellar activity within transmission spectra are difficult to disentangle from low or medium resolution spectra from JWST \citep{Rackham2023Jan}. High resolution spectra that record canonical stellar activity indicators could provide valuable information on risk of stellar contamination within JWST spectra, and could aide greatly in data reduction. Concurrent photometry is useful for monitoring specific events such as spot crossings, but do not provide the insights available from high-resolution spectra of activity indicators such as Ca II H \& K \citep{1978ApJ...226..379W} and H$\alpha$ \citep{Cincunegui2007Jul}. These activity indicators are a way to assess the overall level of activity on the star at the current epoch of observation \citep{Lafarga2021Aug}. The prior H$\alpha$ studies of {\target} have had utility in the past to assess the activity level of the host star, but around a decade has passed between them and JWST Cycle 1 observations. Thus, more recent activity studies on this target are needed to supplement JWST Cycle 1 data to assist in disentangling stellar contamination effects, which is a primary purpose of this paper.

The first results of the JWST observing campaign for HD 189733b were recently published by \citet{Fu2024Jul}. Here they utilized simultaneous photometric monitoring to partially account for effects from unocculted spots. They also observed a star spot crossing during their second observed of two transits. These star spots resulted in a significant shift ($\sim$ 200 ppm) in the transit depth inside the CO$_2$ feature at 4.3-microns. Furthermore, the authors mask out the portion of the transit affected by star spot crossings, which account for about 36\% of their total transit observations. Properly accounting for stellar activity in transmission spectra is crucial to be able to extract as much information as possible out of JWST data.

\section{Observations and Data Reduction}
\indent To examine the H$\alpha$ signal during the transit of {\planet}, we acquired 132 individual spectra from the Multi-Espectrógrafo en GTC de Alta Resolución para Astronomía (MEGARA) spectrograph \citep{dePaz2014Jun} located on the Gran Telescopio Canarias (GTC) 10.4 meter telescope in La Palma, Canary Islands, Spain on 11 June 2022. MEGARA is an optical integral-field Unit (IFU) and multi-object spectrograph (MOS) with a 4000 x 4000 CCD \citep{10.1117/12.925739}.  We used the \lq{}HR-R\rq{} mode of MEGARA, which yields a resolving power of approximately R$\sim$20,000. The HR-R mode spans a wavelength range of approximately 6405.61 \AA - 6797.14 \AA, making it ideal for H$\alpha$ observations at 6562.8 \AA. Our observations were taken with the VPH665-HR grating, which utilizes the IFU mode, and had an exposure time of 11 seconds per frame. The readout time of the instrument is 50 seconds.

The data are recorded in standard \texttt{FITS} file format. We performed standard CCD data reduction steps including subtraction of bias frames, flat fielding, and wavelength calibration using the standard MEGARA data reduction pipeline and the \texttt{numina} package \citep{pascual_2024_14186029}. From here we extracted the 623 x 4300 pixel image, and summed over the rows containing primary fibers to avoid unnecessary exposure to potential cosmic ray contamination. We extracted the single spectral order by summing the rows containing primary fibers (rows 288-290, 292, 306-316, 328, 330, 332-334), selecting the detector rows that encompassed the most signal. This led to a 1D-array of flux that was 4300 pixels long for each frame. Each pixel is in the unit of Jansky. The wavelength dispersion is 0.0974 \AA \hspace{1 pt} per pixel in our configuration. All of our wavelengths listed are air wavelengths.

After reducing the raw frames, we found that 4 of the 132 frames had cosmic rays present in the flux of the H$\alpha$ feature. We discarded these frames, so any irregularities that resulted from cosmic ray correction would not be mistaken for a variation in signal due to stellar activity. Thus, we used 128 total frames for our analysis.

\section{Analysis}
\begin{figure}[h!]
    \includegraphics[width=0.98\columnwidth]{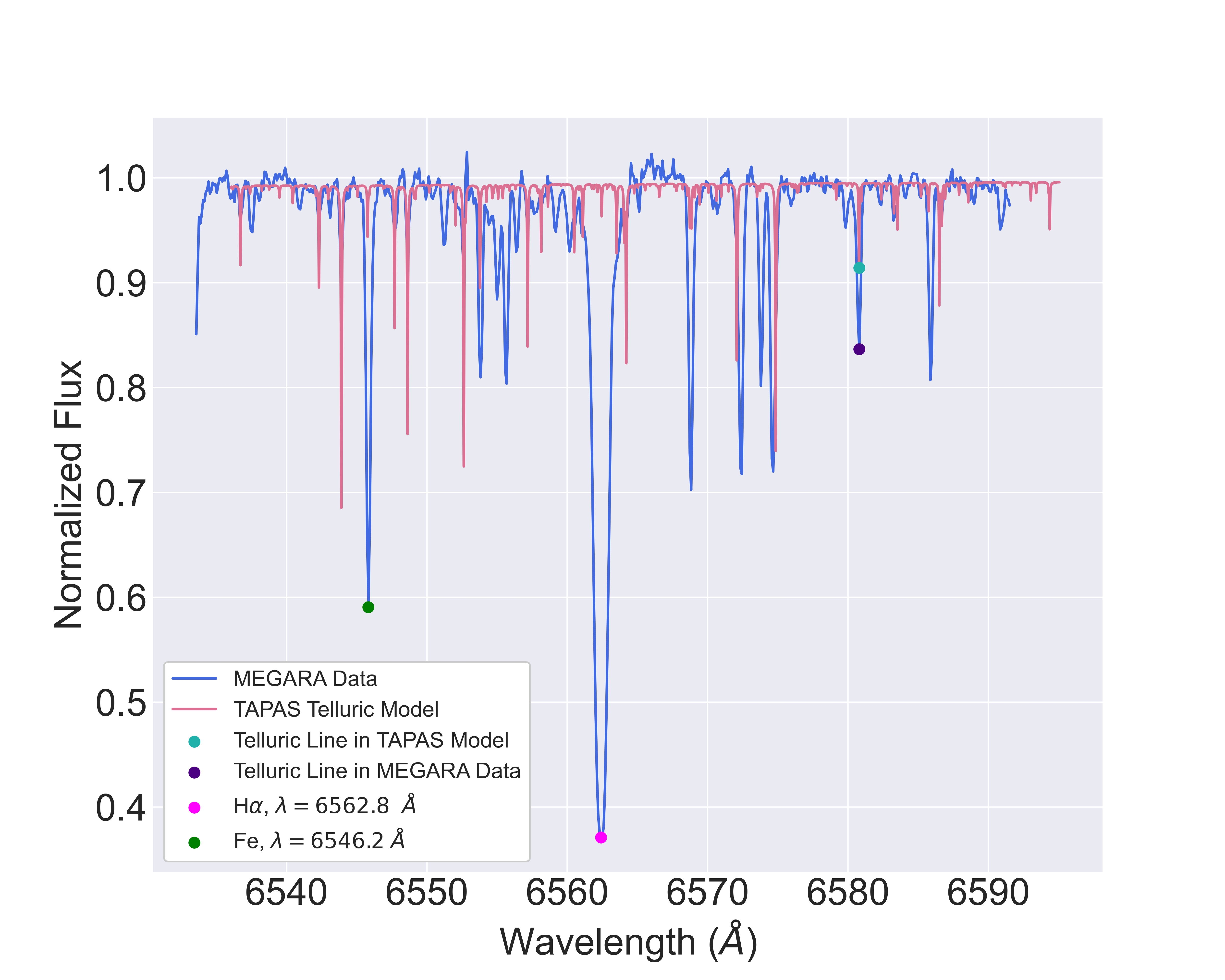}
    \caption{Example R$\sim$20,000 MEGARA spectrum of {\target} after extraction and normalization, with the TAPAS telluric model \citep{Bertaux2014Apr} overlaid in the Earth rest frame. The solid blue line is the MEGARA spectra of HD 189733, and the solid red line is the TAPAS telluric model. The teal dot represents the telluric peak in the MEGARA data, while the violet dot represents the telluric peak in the TAPAS telluric model. Both have a wavelength of $\lambda = 6580.8$ \AA \hspace{2pt}in the Earth rest frame. We also have labeled the H$\alpha$ spectral line in magenta and the Fe I reference line in blue. The Fe I reference line is chosen due to its proximity to H$\alpha$ and its low Land\'e g factor of $g_{eff} = 0.820$, which was retrieved from the Vienna Atomic Line Database 3 (VALD). }
    \label{fig:Selectedline}
\end{figure}

To begin our analysis, we corrected the wavelength range to be in the stellar rest frame. We then narrowed the window of the spectra for our analysis to 6439.06 \AA \hspace{1 pt} to 6692.22 \AA, to focus on H$\alpha$ and the features immediately adjacent to it. To normalize the spectrum in each frame, we used a rolling maximum technique, dividing the spectrum into 26 windows with a wavelength range of 9.35 \AA \hspace{1 pt} each. We then calculated the 99th percentile value for each window, which serves as an estimate of the continuum. Additionally, we rejected the highest point in each flux window to ensure cosmic rays did not interfere with our normalization process. We also removed the window that spanned the width of the broad H$\alpha$ feature. We then fit a spline to interpolate between the 25 continuum points, and divided the spectrum by this function. This results in a spectrum with the continuum normalized to a relative flux value of approximately 1.0, as shown in Figure \ref{fig:Selectedline}.

\hspace{1 pt}We divided the raw summed flux by this spline-fit continuum for the narrow window within each frame to normalize the flux. The raw summed flux is in units of Jansky, but henceforth flux is a unitless quantity due to the normalization.

\begin{figure}[tbh!]
    \includegraphics[width=0.98\columnwidth]{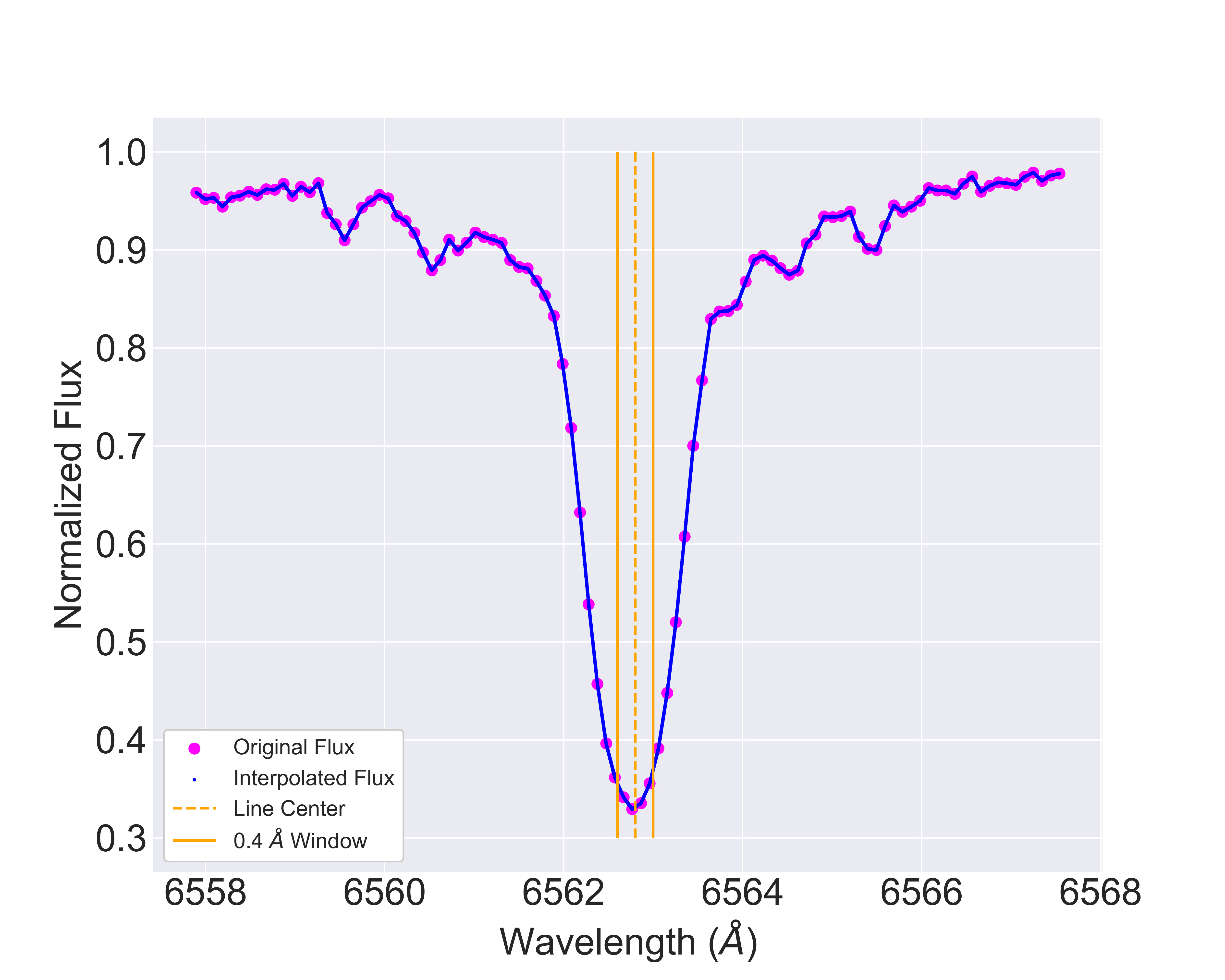}
    \caption{MEGARA spectrum of H$\alpha$ within its feature window of 6557.9 \AA \hspace{1 pt} to 6567.6 \AA. The larger magenta dots represent the original spectral data points, while the smaller blue dots represent the interpolated data points used to integrate along 0.4 \AA \hspace{1 pt}window about the line center (see Section 4). The solid orange lines denote this integration window, and the dashed orange line denotes the line center.}
    \label{fig:linecoreflux}
\end{figure}

\begin{figure*}[t]
    \centering
    \includegraphics[width=0.7\textwidth]{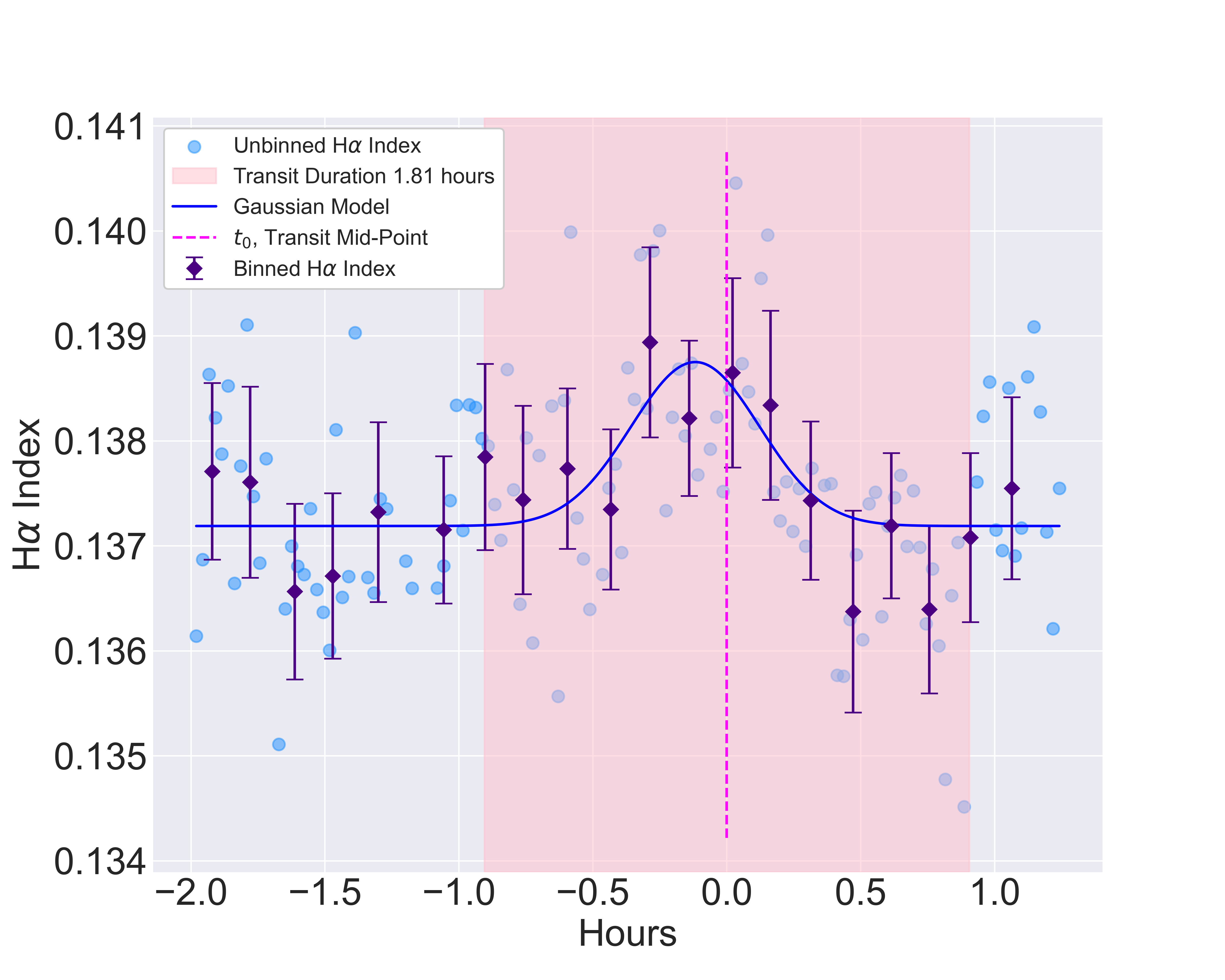}
    \caption{H$\alpha$ index as a function of time in hours, computed using the methodology described in Section 4. The indigo points represent the binned H$\alpha$ indexes with N=6 points per bin, while light blue points represent the unbinned H$\alpha$ indexes. The dashed magenta line shows the transit mid-point and the shaded pink region shows the total transit duration of 1.81 hours. The solid blue line is the Gaussian model that we fit to the data. We measure the increase in H$\alpha$ index by calculating the amplitude of the Gaussian model, which we find to be $\delta = 0.00156 \pm 0.00026$ during the middle of the transit. This is similar in magnitude to the H$\alpha$ index deviations measured by \citealp{2016MNRAS.462.1012B}. This signals that the line is becoming shallower while it is occulted by the planet, after normalizing each frame. We interpret this to be a star spot crossing, based on the results of our simulated H$\alpha$ transit shown in Figure~\ref{fig:smart} and Figure~\ref{fig:enter-label}. This deviation is consistent with the planet crossing a star spot with a radius of $= 3.47 \pm 0.30 R_{\oplus}$. Additionally, we measure the full-width half-max (FWHM) of the Gaussian model of the H$\alpha$ indexes to be 0.57 $\pm$ 0.13 hours (34 $\pm$ 8 minutes), which is comparable to the time it would take for the planet to cross itself, about 24 minutes. This is in line with an occultation of an active region that is smaller than the planet itself, since the planet's diameter then drives the occultation timescale.}
    \label{fig:Halphadepths}
\end{figure*}

In this work we define the H$\alpha$ index by numerically integrating using \verb|scipy.integrate.trapz| along a 0.4 \AA \hspace{1 pt} window centered on the line minimum, as illustrated in Figure~\ref{fig:linecoreflux}. This is similar to the H$\alpha$ index defined in \citealp{2016MNRAS.462.1012B}. We use this index to determine the change in occulted active regions along the transit chord as {\planet} transits, with baseline time before and after the transit. We repeat this reduction for each frame, and show our resultant time series in  Figure~\ref{fig:Halphadepths}. Figure~\ref{fig:Halphadepths} shows an increase in the value of the H$\alpha$ index, which is a decrease in depth, peaking around the mid-transit time. Compared to the pre-transit baseline, we observe an increase of our H$\alpha$ index of $\delta = 0.00156 \pm 0.00026$ compared to the median out-of-transit flux. Due to observational scheduling constraints, there are minimal post-transit observations and so the behavior of the H$\alpha$ index is more uncertain after the transit egress. 

We calculated the uncertainties on flux indexes using a bootstrapping method. We identified a quiet section of the continuum (between 6490 and 6491 \AA) and took the standard deviation of the 10 flux data points in this continuum section prior to normalization. For each data point in the H$\alpha$ and Fe I absorption lines, we resample each data point by adding random Gaussian noise based on the measured continuum standard deviation multiplied by the square root of the ratio of the individual point flux over the continuum flux, since Poisson-like noise approximately follows $\sigma = \sqrt{F}$. Thus, the random Gaussian noise per point can be described by $\sigma_{p} = \sigma_{c}\sqrt{(F_{p}/F_{c})}$, where $c$ refers to the continuum and $p$ refers to the individual data points within the spectral features. We then re-interpolated and integrated based on the resampled data points.  We repeated this procedure 1000 times per frame, and took the standard deviation of the resultant integrated H$\alpha$ indexes to estimate the uncertainty. We add the uncertainties derived from this bootstrapping method for each frame in quadrature within each bin and divide by $\sqrt{6}$, since there are 6 data points per bin. 

As tests to see whether the trend we observe in Figure~\ref{fig:Halphadepths} might be associated with stellar activity, we compared the H$\alpha$ indexes to an activity-insensitive feature as a baseline, to airmass throughout the night, and to a nearby telluric line to ensure there was no correlation with any of these possible noise sources. To identify a non-active line for comparison, we began by extracting stellar features using the Vienna Atomic Line Database 3 (VALD) (\citealp{2019ARep...63.1010P}, \citealp{BPM}, \citealp{BWL}, \citealp{K14}). We extracted stellar features among the previously mentioned narrowed spectra window of 6531.6 \AA \hspace{1 pt}to 6591.9 \AA \hspace{1 pt}using stellar parameters of {\target} from \citealp{10.1093/mnras/stu2502}. We then identified specific transitions in our observed spectra by matching them to lines from VALD, and selected the deepest line, which we identified to be Fe I ($\lambda = 6546.2$ \AA), as seen in Figure~\ref{fig:Selectedline}. We verified that this line is not known to be sensitive to stellar activity by checking \cite{Wise_2018}, which probes thousands of lines from 2 K-star spectra and determines 40 new activity indicators. This iron line was not among them, and was not found to be magnetically sensitive. This line also has a Land\'e G factor of 0.820, which is less than $g_{eff} = 1.2$, which is considered the threshold below which a line is magnetically insensitive \citep{Bellotti2022Jan}.

\begin{figure}[h!]
    \includegraphics[width=0.98\columnwidth]{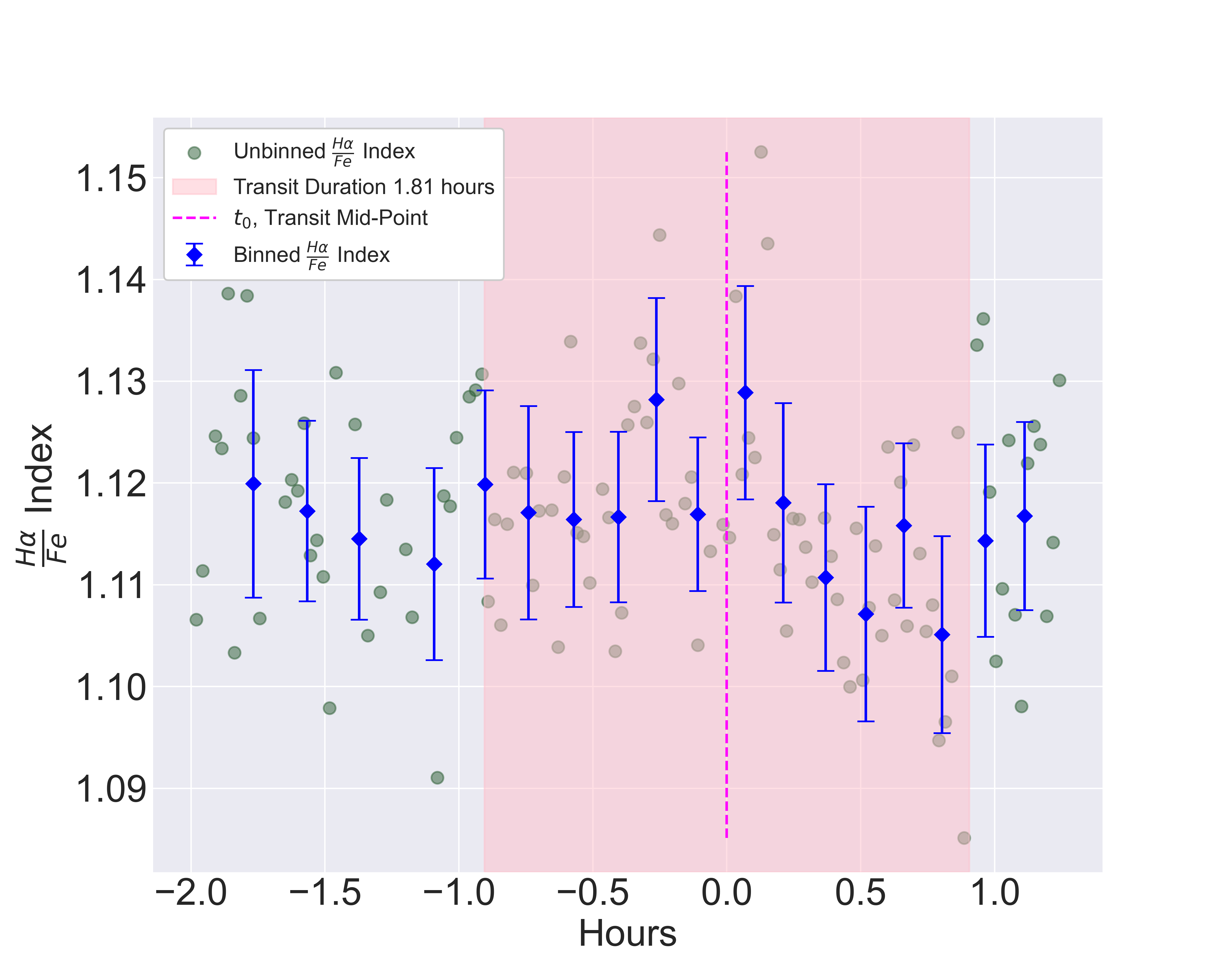}
    \caption{H$\alpha$ index divided by Fe I index. The blue dots represent the binned index ratio, while the pine green dots represent the unbinned index ratio. The pink shaded region shows the transit duration, and the dashed magenta line shows transit mid-point. This figure shows similar behavior to Figure~\ref{fig:Halphadepths}, which indicates that the source of the signal we observe in the H$\alpha$ index time series is likely not significantly affecting the control Fe I line we observe here. We calculated a Pearson Correlation Coefficient between H$\alpha$ index and Fe I index and find a coefficient of $\rho_{H\alpha,Fe I} = -0.0051$ with a p-value of $p = 0.95$, showing there is no statistically significant correlation in Figure~\ref{fig:HalphaFeIcorr}.}
    \label{fig:HalphaFeI}
\end{figure}
\begin{figure}[]
    \includegraphics[width=0.98\columnwidth]{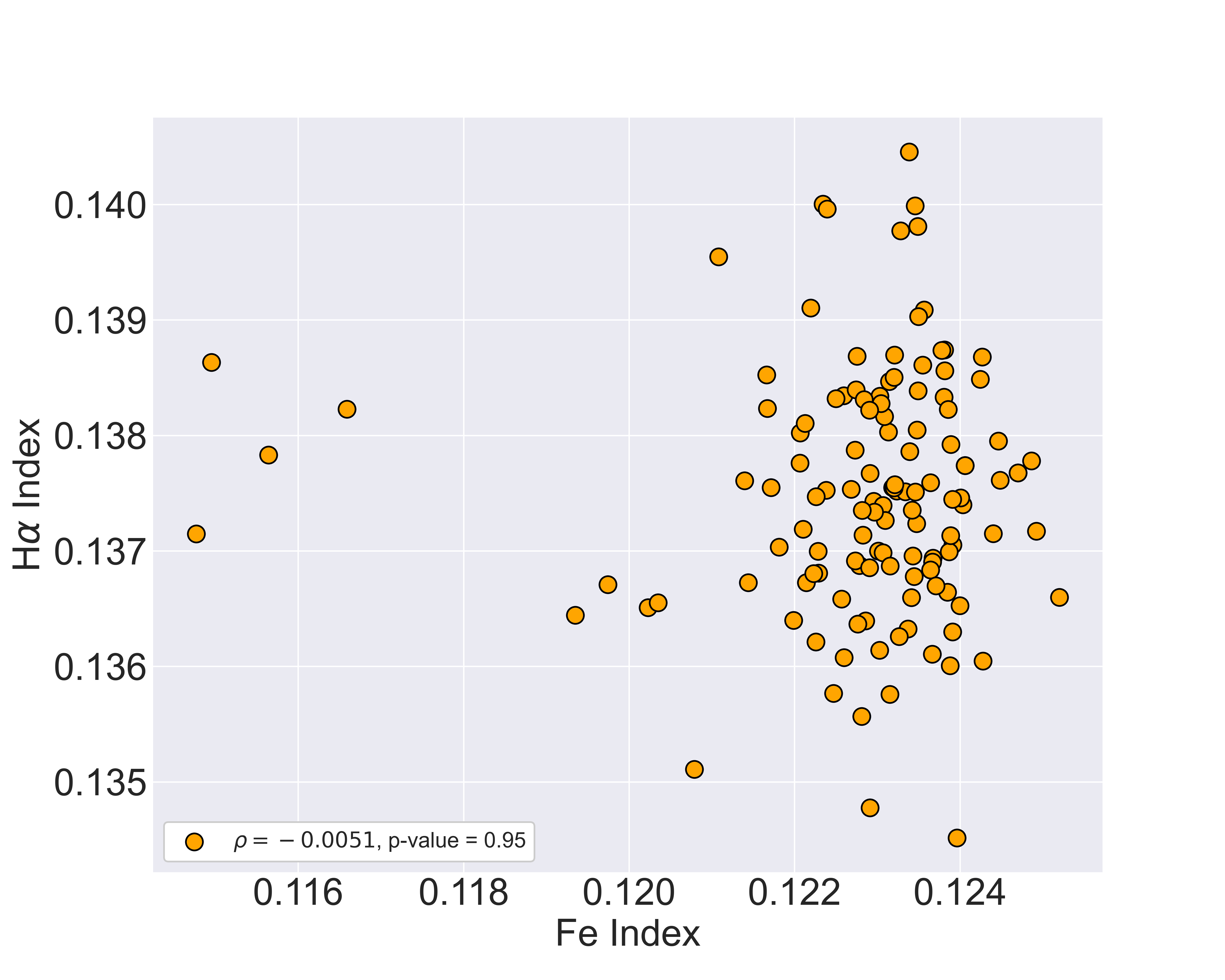}
    \includegraphics[width=0.98\columnwidth]{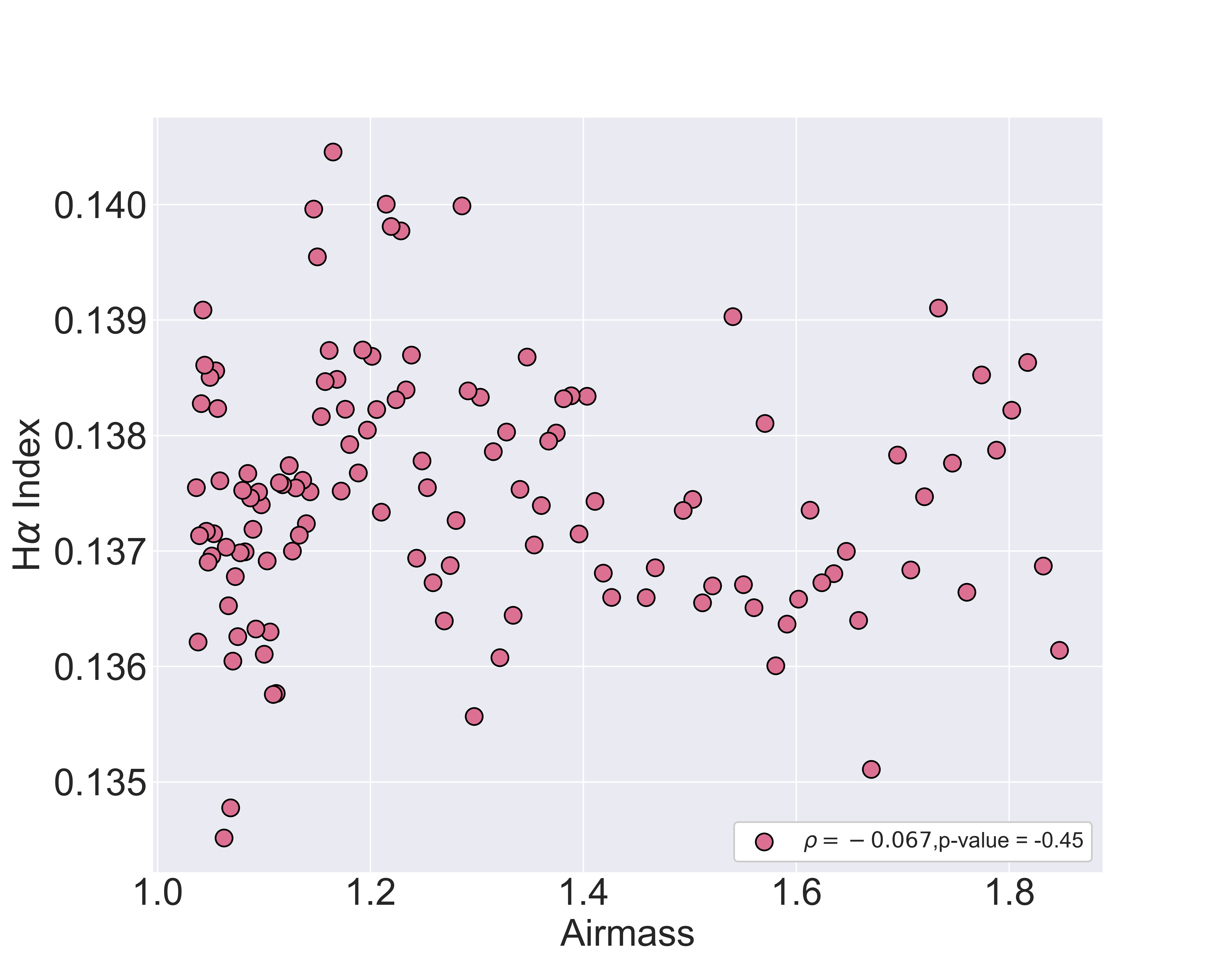}
    \caption{Correlation plots of two methods used to check that H$\alpha$ signal is stellar in origin. Top: H$\alpha$ index as a function of Fe I index. We measure a Pearson R Correlation Coefficient of $\rho_{H\alpha,Fe\hspace{1pt}I} = -0.0051$ with a p-value of $p = 0.95$ between H$\alpha$ index and Fe I index, and $\rho_{H\alpha, airmass} = -0.067$ with a p-value of $p = 0.45$ between H$\alpha$ index and airmass. No statistically significant correlation is present, which suggests that the origin of the signal variation we observe in H$\alpha$ does not appear to strongly impact Fe I. Since this Fe I line is not magnetically sensitive, the origin of this H$\alpha$ signal could indeed be stellar activity. Bottom: H$\alpha$ index as a function of airmass of the observation. We again do not see any statistically significant correlation, which suggests that airmass or other observational effects are not the origin of this H$\alpha$ variation either.}
    \label{fig:HalphaFeIcorr}
\end{figure}

We then measured a comparable index (also a 0.4 \AA \hspace{1 pt} window) for this Fe I line, which we refer to as the baseline feature. If the signal we are seeing in the time series of the H$\alpha$ index is truly originating from magnetic activity, and the baseline feature is not sensitive to this activity, then we should see a similar trend when dividing the H$\alpha$ index by the Fe I index. Indeed, we do see a similar trend in Figure~\ref{fig:HalphaFeI}. We also verified that the time series of Fe I indexes did not have any artifacts present that could influence the shape seen in Figure~\ref{fig:HalphaFeI}, and that it appeared to be uncorrelated scatter. We additionally calculated a Pearson Correlation Coefficient between H$\alpha$ index and Fe I index and found a coefficient of $\rho_{H\alpha,Fe I} = -0.0051$ with a p-value of $p = 0.95$, showing there is no statistically significant correlation in Figure~\ref{fig:HalphaFeIcorr}. This suggests that the H$\alpha$ variation is due physically to something that is not altering Fe I index.

Next, we verified the trend seen in Figure~\ref{fig:Halphadepths} was not due purely to correlations with the airmass of the observations throughout the night. The airmasses when our spectra were taken were all relatively low ($<$ 1.8), and there did not appear to be any correlation present between H$\alpha$ and airmass. We verified this with a Pearson Correlation Coefficient, which had a value of $\rho_{H\alpha, airmass} = -0.067$ and a p-value of $p = 0.45$ between , showing there is no statistically significant correlation and we have plotted our H$\alpha$ index vs the airmass of the observation in  Figure~\ref{fig:HalphaFeIcorr}. Although airmass is related to time, it decreases throughout the night, so the signal we observe in H$\alpha$ does not appear to be an airmass-driven effect.

Our final test to determine if our measured H$\alpha$ variation is correlated with terrestrial atmospheric effects was to compare our observations with telluric lines imprinted in our spectra. To do this, we used the TAPAS web service \citep{Bertaux2014Apr} to get a site-specific telluric model. We shifted our MEGARA data back into the Earth frame to match the telluric model's rest frame. Both the MEGARA and TAPAS model wavelengths are air wavelengths rather than vacuum, so no adjustments were needed in that regard. After this, we overplotted the model on top of our spectra (see Figure \ref{fig:Selectedline}).

\begin{figure}[]
    \includegraphics[width=0.98\columnwidth]{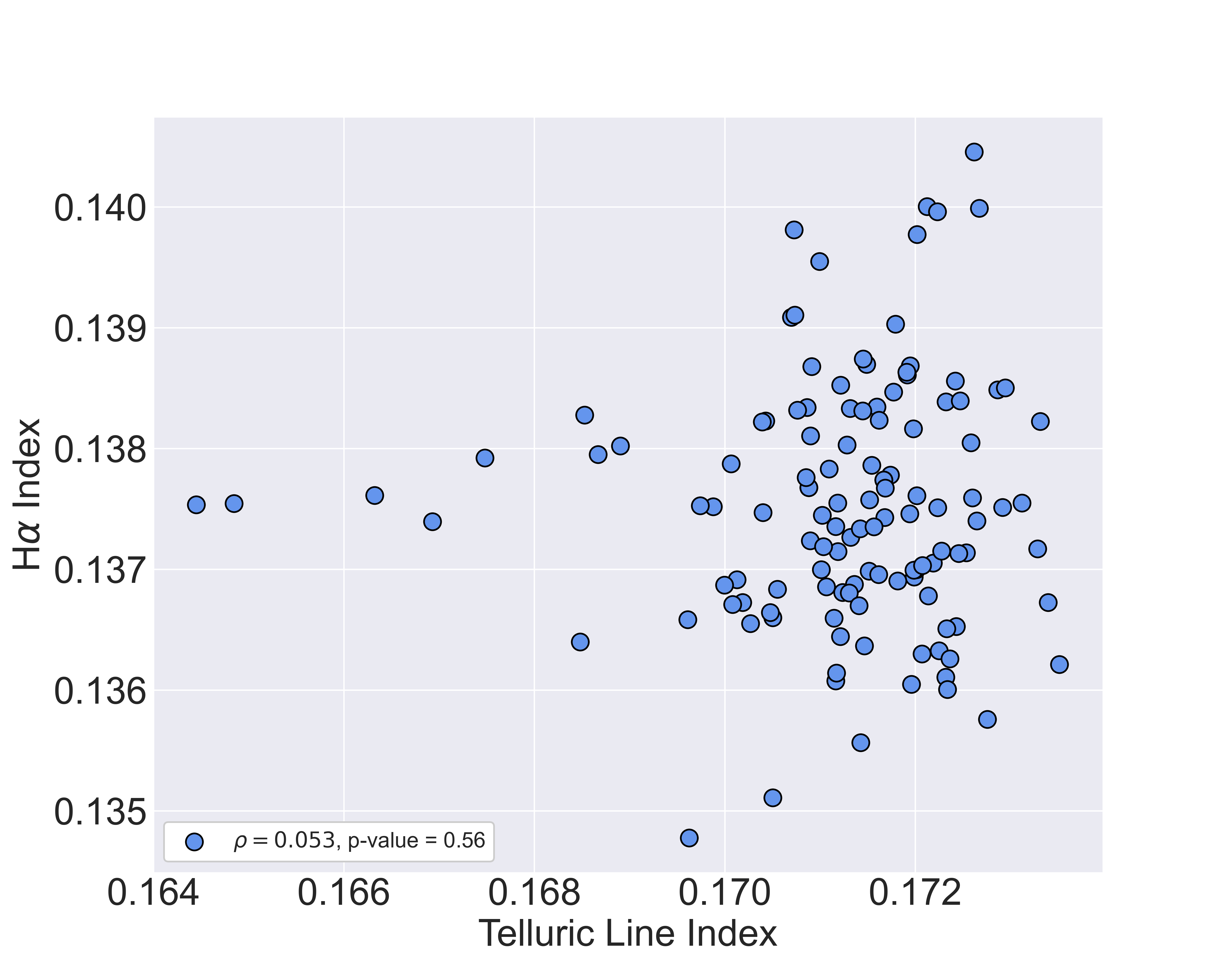}
    \caption{Correlation plot of H$\alpha$ Index versus Index for Telluric Line ($\lambda = 6580.8$ \AA) selected from the TAPAS Telluric model. This shows the data after removing the 4 outlier points to better illustrate the scatter. We find a Pearson Correlation Coefficient of $\rho_{H\alpha, T} = 0.053$ with a p-value of $p = 0.56$ excluding outlier points, which shows there is no statistically significant correlation present.}
    \label{fig:tellcorr}
\end{figure}

We then employed \texttt{find\_peaks} again to find the wavelengths of the TAPAS telluric features, and match them with features in our data. We selected a telluric line at $ \lambda = 6580.8$ \AA \hspace{1 pt} since it aligned best with our data within the narrowed spectra wavelength window. Figure~\ref{fig:Selectedline} shows the TAPAS telluric model overplotted against our MEGARA data, with the selected telluric feature. We then measured the telluric line index, integrated in a 0.1 \AA window due to the narrow and shallow nature of the line in each frame, and again checked for correlation with H$\alpha$ and found none. We calculated a Pearson Correlation Coefficient between H$\alpha$ and telluric indexes, and find there is no statistically significant correlation. We removed four outlier points which were likely a product of continuum normalization, and calculated a Pearson Correlation Coefficient before and after removing the outlier points. Including outlier points, the Pearson Correlation Coefficient value for H$\alpha$ versus telluric indexes was $\rho_{H\alpha, T} = 0.209$, with a p-value of $p = 0.018$. After removing outlier points, we find a Pearson Correlation Coefficient of $\rho_{H\alpha, T} = 0.053$ with a p-value of $p = 0.56$. This is further evidence that the patterns we observe are not likely a result of telluric contamination and variation. This is shown in Figure~\ref{fig:tellcorr}.

In light of these tests, we suggest that the origin behind the observed variation in H$\alpha$ is likely star spots, since unocculted spots can strengthen absorption features \citep{Thompson_2024}. Therefore, since we observe a variation in H$\alpha$ during the transit, it is probable that there are spots along the transit chord that are being obscured by the transiting planet, which causes the depth of H$\alpha$ to decrease. After performing the checks on H$\alpha$, we computed the how large of an active region this change in depth of H$\alpha$ corresponds to. We use Formula~\ref{spoteq} to solve for 
 $R_*$. This is a formula typically used to measure transit depth, so we use it to estimate the radius of the active region the planet is occulting.

\begin{equation} \label{spoteq}
\delta = \frac{R_{spot}^2}{R_*^2}
\end{equation}
    
\begin{equation} \label{spotr}
R_{spot} = (R_*^2 \delta )^{1/2}
\end{equation}

Where $\delta$ is the amplitude of the Gaussian model fit to the unbinned H$\alpha$ indexes, $R_*$ is the radius of HD 189733, and $R_{spot}$ is the estimated star spot radius. We fit a Gaussian model to the unbinned data to estimate the amplitude and temporal length of the spot-crossing event. We measure the amplitude of the Gaussian to be $\delta = 0.00156 \pm 0.00026$ with a FWHM = 0.57 $\pm$ 0.13 hours. We calculate the uncertainties on the parameters of the Gaussian model by resampling each unbinned H$\alpha$ index data point with Gaussian noise with a standard deviation equal to its associated uncertainty. We then fit a Gaussian model to the resampled data and measure each parameter. We perform this procedure 1000 times, and take the standard deviation of each parameter's distribution to be its corresponding uncertainty. In addition to the Gaussian model, we tested a flat line with an offset. The data favored the Gaussian model, which had a reduced chi-squared value closer to 1 than the flat line model.

Using Equation~\ref{spotr}, we estimate the deviation is caused by a relatively small spot (or spot-cluster) radius of $R_{spot} \approx 3.47 \pm 0.30 R_{\oplus}$. We assume a stellar radius of $0.806 \pm 0.016 R_{\odot}$ (\citealp{Boyajian2015Feb}, \citealp{Cegla2016Apr}).

\section{Discussion}
\label{sec:discussion}
Proper analysis of transmission spectra often requires an understanding of the relative level of stellar activity along the transit chord at the time of transit, to either make corrections in the data or to omit data from analysis (\citealp{10.1093/mnras/staf489}, \citealp{article}, \citealp{Fu2024Jul}). The most recent H$\alpha$ study of {\target} was published in 2016 \citep{2016MNRAS.462.1012B}, so our spectra were taken much closer to the relevant epoch of JWST Cycle 1 observations of HD 189733b. We aim to determine how active the transit chord, and in turn the entire star, is and to assess the risk of stellar contamination in its transmission spectra. Based on our analysis of H$\alpha$, we find it likely that the star was active with stellar inhomogeneities present on its surface close in time to when the JWST observations were taken. This means that the transmission spectra are at risk for stellar contamination, and spurious molecular features may be present or enhanced. There is evidence for stellar contamination in prior transmission spectra taken of HD 189733b \citep{2016MNRAS.462.1012B}. This can be mitigated to an extent by modeling stellar activity using techniques such as Gaussian processes \citep{10.1093/mnras/stac1949}. However, it is important to be aware of the level of stellar activity present to distinguish between genuine planetary features and signals of stellar origin in an inferred transmission spectrum.

Prior analysis to determine stellar activity levels of HD 189733b using H$\alpha$ transits has been performed by \citealp{2016MNRAS.462.1012B} and \citealp{Cauley2017Apr}. These analyses were performed nearly a decade prior to JWST observations being taken. With this in mind, in the intervening decade, HD 189733 may be at a different part of its stellar cycle and additional observations are warranted.

\citeauthor{Cauley2017Apr} used H$\alpha$ equivalent width measurements and found that H$\alpha$ did not always correlate strongly with the Ca II H \& K, but that the largest H$\alpha$ absorption occurred when the star was most active. \citeauthor{2016MNRAS.462.1012B} use a similar definition of H$\alpha$ index to that defined in this work, but with a line core flux window of 0.7 \AA \hspace{1pt}, slightly larger than our window of 0.4 \AA \hspace{1pt}. They measure H$\alpha$ index deviations comparable to ours, detecting excess H$\alpha$ absorption of $\delta = 0.0074 \pm 0.0044$ and $\delta = 0.0214 \pm 0.0022$ during their first and second transits, respectively. This is similar in magnitude to the deviation we measure of $\delta = 0.00156 \pm 0.00026$. Based on this, it is likely that HD 189733 is experiencing a similar level of stellar activity to that observed in previous studies from the decade prior. Therefore, stellar contamination in HD 189733b's transmission spectrum is likely an issue that must be accounted for in Cycle 1 JWST observations.

\begin{figure*}[t]
    \centering
    \includegraphics[width=0.7\textwidth]{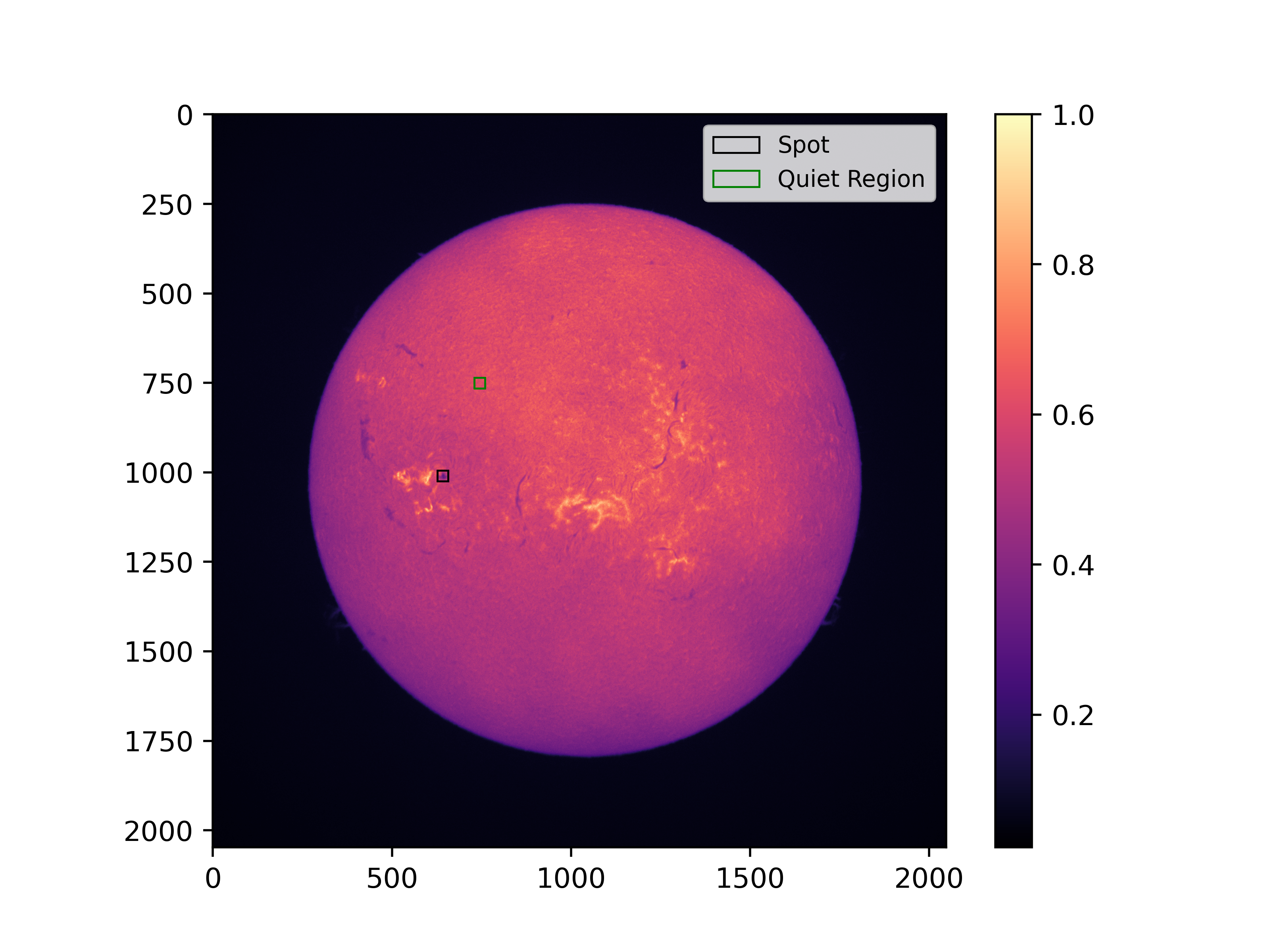}
    \caption{SMART image of H$\alpha$ line center of the solar disk, taken on July 12, 2016 featuring AR 12565. The black box is centered around the active region, and the green box is centered around a quiet region from which flux is pulled for comparison. The image is normalized by the maximum solar flux value. We find $\mathrm{Flux}_{quiet} \sim 0.61$ and $\mathrm{Flux}_{spot} \sim 0.37$, showing that spots deepen absorption in the H$\alpha$ line core.}
    \label{fig:smart}
\end{figure*}

We postulate that the cause of the variation in H$\alpha$ is a spot or group of spots being obscured along the transit chord since we observe H$\alpha$ becoming more shallow during transit. Unocculted star spots are known to strengthen absorption features \citep{Thompson_2024}, so a potential scenario is that the planet is obstructing a spot or group of spots along that transit chord that is contributing to the total H$\alpha$ absorption we observe in the unocculted stellar disk. When the spot is unocculted along the transit chord, it deepens the H$\alpha$ absorption signal, and when the planet occults the spot, the H$\alpha$ absorption feature becomes shallower until it is unocculted again. Although \citealp{Thompson_2024} points out that unocculted spots can deepen absorption features, spot effects on activity indicators are more complicated due to the magnetic sensitivity of the spectral lines and atmospheric dynamics. To test the effects a spot would have on the H$\alpha$ index, we utilize archival data from the Solar Magnetic Activity Research Telescope (SMART) at Hida Observatory\footnote{https://www.hida.kyoto-u.ac.jp/SMART/}, which takes spatially resolved full-disk spectroscopic images of the Sun in H$\alpha$.  We pull an image taken in the H$\alpha$ line center, and the filter has a FWHM of 0.25 $\mathrm{\AA}$, which is comparable to our index width of 0.4 $\mathrm{\AA}$. We use an image taken on July 12, 2016, which features Active Region (AR) 12565. This image is shown in Figure \ref{fig:smart}. We normalized the frame by dividing by the maximum flux value in the Sun.  There is an isolated spot present, so we measure H$\alpha$ line core flux in the spot, and on a uniform area of the solar surface to serve as a quiet activity proxy. We do indeed see that the H$\alpha$ line core flux is lower in the spotted area compared to the quiet area, with the selected quiet area having a normalized flux value of $\mathrm{Flux}_{quiet} \sim 0.61$ and the spot flux measuring to be $\mathrm{Flux}_{spot} \sim 0.37$. This indicates that H$\alpha$ absorption is indeed deeper in star spots. This aligns with our interpretation, since H$\alpha$ absorption becomes shallower as the planet occults the star (and thus spot) mid-transit.

\begin{figure}
    \centering
    \includegraphics[width=0.98\linewidth]{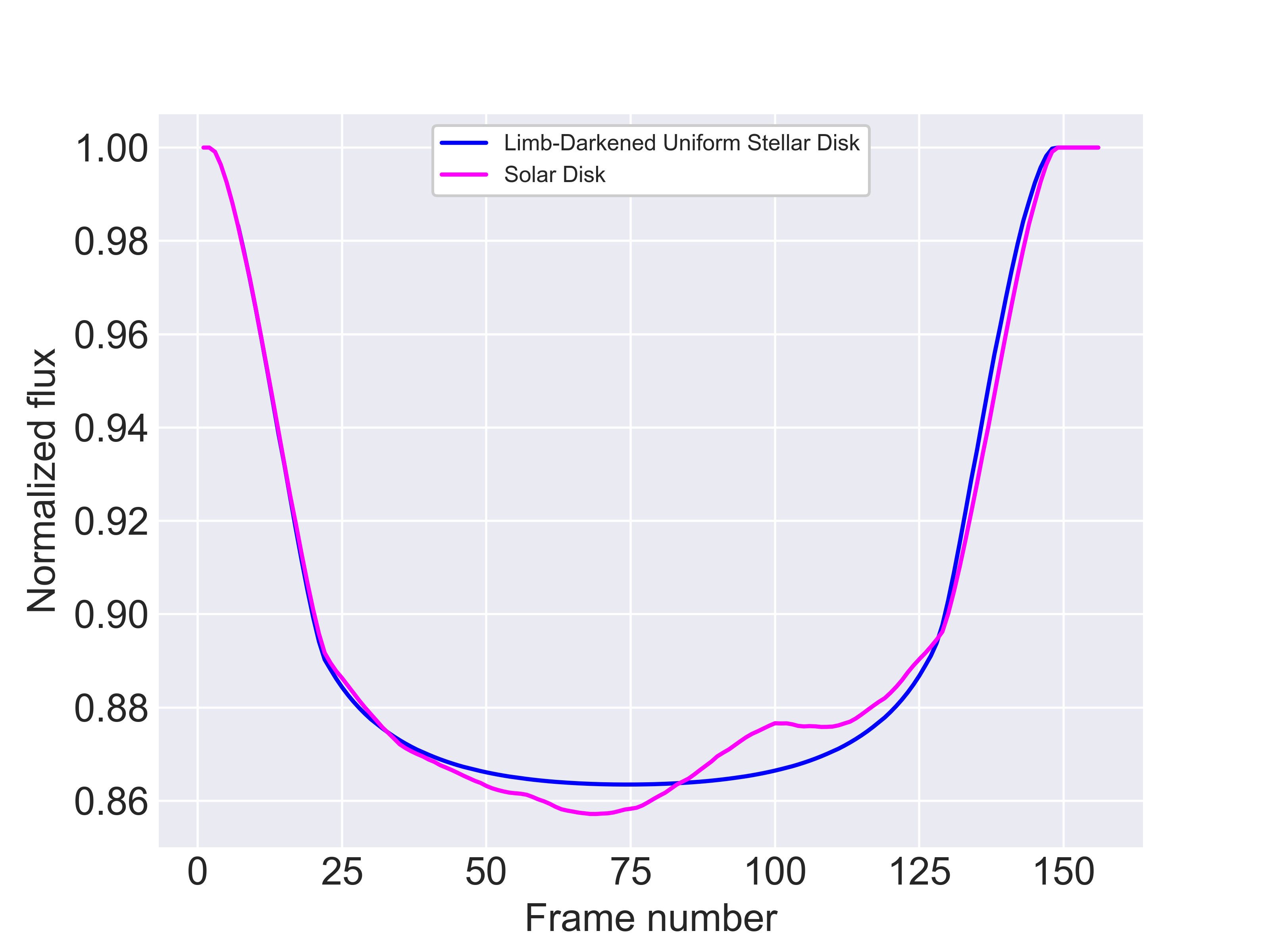}
    \caption{Simulated transit of H$\alpha$ data from Figure~\ref{fig:smart}. The magenta line shows a transit across the actual solar disk, with an impact parameter of 0 to make the transit chord span the most active regions of the solar disk. The blue line shows a transit across a simulated uniform stellar disk that has been limb darkened using a quadratic law, with relevant coefficients of a = 0.2 and b = 0.5 obtained from \citealp{Neilson2013Jun}.}
    \label{fig:enter-label}
\end{figure}

To ensure we are viewing a variation due to the crossing of an active region and not a geometric effect, we searched the literature to verify that center-to-limb variation (CLV) and the Rossiter-McLaughlin (RM) effects are low for H$\alpha$ specifically, since these effects are line-dependent due to their different formation depths. CLV and RM effects are indeed weak for the H$\alpha$ core, with a variation in flux $F_{it}/F_{oot} -1$ of $\sim 0.1 \%$ \citep{Pietrow2023Mar}. We observe a contrast of 2.4 $\%$ within our data, which is over an order of magnitude larger than potential geometric effects. To further test this, we simulated a fake transit across the aforementioned Hida Observatory SMART solar $H\alpha$ core image with the same $R_p/R_*$ ratio as HD 189733b. We simulate a transit across the solar disk, and simulate a uniform, limb-darkened disk using the quadratic limb-darkening law with relevant coefficients obtained from \citealp{Neilson2013Jun}. To simulate the transit, we mask out a circle of pixels the size of our fake planet and sequentially "stamp" it across the original H$\alpha$ image. Each time the fake planet mask is stamped across the image, we sum the total H$\alpha$ flux in the solar disk. We then normalize the overall light curve with the out of transit baseline flux value. We see that effects from stellar activity and heterogeneities dominate, causing the transit to clearly deviate from the model that shows pure CLV (Figure~\ref{fig:enter-label}).

\citealp{Fu2024Jul} are the first to publish their analysis of JWST NIRCam spectra (R $\sim$ 1,600) for HD 189733b, as part of the Cycle 1 GO program. They report detections of H$_2$O, CO$_2$, CO, and H$_2$S in its transmission spectrum, which they observed on 25 August and 29 August 2022, a few months and approximately 6.77 stellar rotation cycles after our observations. They model for stellar heterogeneities using equations (4) and (5) from \citealp{2011MNRAS.416.1443S} assuming a dimming from unocculted spots of $\sim$2.7 percent and $\sim$1 percent for the first and second visit, respectively, from ground-based photometry. %While they model broadly for stellar heterogeneity, disentangling specific effects is not possible from photometry.
Based on the duration of their recorded spot crossing, we estimate the spot they observed to have an approximate radius of $\sim$ 3.6 Jupiter radii, which is about 8 times larger than the spot we record along the transit chord in this work. Their data were taken 79.84 days after our data, so assuming the same rotational period they recovered from photometric monitoring of about 11.8 days, we observe a different stellar disk than they do. Therefore we suggest the spot we observe is unlikely to be the same spot they observe, despite the fact that spot lifetimes would make that a possibility \citep{Giles2017Dec}. 

Based on how \citealp{Fu2024Jul} model unocculted spot and stellar contamination effects their transmission spectrum, we estimate that a spot similar in size to the one we observe ($R_{spot} = 3.47 R_{\oplus}$) could shift the transit depth up by approximately 17 ppm, in the 4.2 to 4.7 micron CO$_2$ feature. This is also assuming a ~1 percent dimming from unocculted features, which is taken from \citealp{Fu2024Jul}. Along with unocculted spot dimming, their recorded spot shifted the transit depth of the same wavelength regime up by approximately 0.0002, or 200 ppm. This demonstrates that even a smaller spot comparable to the one we observe could have significant ramifications for JWST results, since the NIRCam Grism modes have a precision of $\sim$ 20 ppm \citep{Rieke2023Feb}. Thus, a spot similar in size to ours ($R_{spot} = 3.47 \pm 0.30 R_{\oplus}$) could affect a transmission spectrum in a non-negligible way. This effect could be more subtle for single transits, but its impacts could become exacerbated in instances where multiple transits are averaged together \citep{article}. Such subtle fluctuations can influence molecule detection confidence intervals and atmospheric retrievals, biasing inferred results, and can be difficult to detect. While the larger spot crossings such as the one observed by JWST \citep{Fu2024Jul} are obvious in concurrent photometric observations, smaller spots such as the one we detect are more challenging to account for since they lie closer to the JWST noise floor. 

We note that the field's current understanding of JWST data products and best practices regarding their reduction are still in their early stages and there are likely to be improvements over the course of the JWST mission. However, the current procedure for analyzing exoplanet data products affected by stellar activity (spots, as well as flares) is to mask out this data and exclude it from analysis completely. \citealp{Fu2024Jul} corrects for the recorded spot occultation in one of their two transits by masking out the contaminated data during the transit, which is approximately $\sim$ 36$\%$ of their total in-transit data. Additionally, masking out occulted active regions can still leave residual spot effects imprinted onto the inferred planetary transmission spectrum, such as a slope (\citealp{Fournier-Tondreau2024Feb}, \citealp{Bixel2019Feb}, \citealp{Oshagh2014Aug}). More advanced methods to correct these data rather than discard them can increase the total measured signal in understanding the planet's atmosphere. High-resolution spectral monitoring of the host star during these observations may be a promising future avenue to investigate new ways to correct for stellar activity in JWST data. We recommend simultaneous high-resolution spectroscopic observations of activity indicators concurrent with JWST exoplanet observations whenever possible in order to obtain a fuller understanding of the activity level present on the host star.
High-resolution spectra can reveal more subtle fluctuations of the activity level present than photometry can, particularly since different activity indicators trace different types of activity, such as photospheric versus chromospheric activity. Atomic properties of activity indicators offer insight to where they are formed and what kind of activity they trace. For example, spectral lines with low excitation energies probe the upper photosphere, whereas lines with low excitation energies probe the lower photosphere \citep{Solana2005Aug}. Identifying the kinds of activity present and their origins and disentangling their effects from spectra may be key to utilizing the power JWST has to offer with respect to exoplanetary atmospheric measurements.

\section{Conclusion}
HD 189733b is an ideal candidate for characterization due to its bright nature and easily observable signal, however, stellar contamination must be considered as the host star is known to be active. Previous H$\alpha$ activity studies have been performed on {\target} (\cite{2016MNRAS.462.1012B}, \cite{Cauley2017Apr}), but they are not as informative of the current epoch since the most recent publication was in 2016. JWST low-resolution spectroscopy offers potential to characterize this planet, but would benefit from additional considerations and modeling to assist in disentangling stellar and planetary signals. We utilize high-resolution spectra from the MEGARA spectrograph on GTC to measure variations in the shape of H$\alpha$ during a transit of HD 189733b. We measure a change in H$\alpha$ index of $\delta = 0.00156 \pm 0.00026$, and deduce that the likely cause of this variation is the planet temporarily occulting a star spot along the transit chord with a radius of $R_{spot} = 3.47 \pm 0.30 R_{\oplus}$. The spot's contribution to the H$\alpha$ line is obscured for part of the transit, making the line appear more shallow while the spot is occulted during the transit. Simultaneous photometric monitoring is useful for larger features, but is less sensitive to smaller features that are closer to the noise floor of the instrument which can still impact important measurements of detections within transmission spectra and atmospheric retrievals. 

\section{Acknowledgments}
We thank the referee for their comments and suggestions, which improved the quality of this work. This work is based on data obtained with MEGARA/MIRADAS instrument, funded by European Regional Development Funds (ERDF), through Programa Operativo Canarias FEDER 2014-2020. Based on observations made with the Gran Telescopio Canarias (GTC), installed at the Spanish Observatorio del Roque de los Muchachos of the Instituto de Astrofísica de Canarias, on the island of La Palma. The research was carried out, in part, at the Jet Propulsion Laboratory, California Institute of Technology, under a contract with the National Aeronautics and Space Administration (80NM0018D0004). This work has made use of the VALD database, operated at Uppsala University, the Institute of Astronomy RAS in Moscow, and the University of Vienna. We express our sincere gratitude to the staff of the
Hida Observatory for the development and maintenance of the
instrument and daily observation.

\begin{deluxetable*}{lcccccc}
\tabletypesize{\scriptsize}
\tablewidth{0pt}
\tablecaption{MEGARA Dataset (First 10 rows) \label{table} }
\tablehead{\colhead{Time from $t_0$(hrs)} & \colhead{H$\alpha$ Index} & \colhead{H$\alpha$ Index Error} & \colhead{Fe I Index} & \colhead{Fe I Index Error} & 
\colhead{Airmass}}
\startdata
% \multicolumn{6}{c}{} \\
\hline
-1.980	& 0.13614	& 0.00073 & 	0.12303	& 0.00077 &	1.85 \\
-1.957 &	0.13687	& 0.00058 &	0.12316	& 0.00062	& 1.83 \\
-1.933	& 0.13863	& 0.00010	& 0.11495	& 0.0010 &	1.82 \\
-1.909 &	0.13822	& 0.00083	& 0.12291	& 0.00083 & 1.80 \\
-1.886 &	0.13787	& 0.00044	& 0.12273	 & 0.00046 &	1.79 \\
-1.862 & 	0.13852	& 0.00064	& 0.12166	& 0.00065 & 1.77 \\
-1.838	& 0.13664	& 0.00097	& 0.12385	& 0.00095 &	1.76 \\
-1.815 &	0.13776	& 0.00083	& 0.12207	& 0.00079 &	1.75 \\
-1.791 &	0.13910	& 0.0011	& 0.12219	& 0.0011 &	1.73 \\
-1.767	& 0.13747	& 0.00050	& 0.12226	& 0.00053	& 1.72 \\
-1.744 & 	0.13684	& 0.0010	& 0.12364	& 0.0011 &	1.71 \\
\hline
\enddata
\end{deluxetable*}

\bibliographystyle{aasjournalmod}
\bibliography{bibliography.bib}

\begin{thebibliography}{}
\expandafter\ifx\csname natexlab\endcsname\relax\def\natexlab#1{#1}\fi
\providecommand{\url}[1]{\href{#1}{#1}}
\providecommand{\dodoi}[1]{doi:~\href{http://doi.org/#1}{\nolinkurl{#1}}}
\providecommand{\doeprint}[1]{\href{http://ascl.net/#1}{\nolinkurl{http://ascl.net/#1}}}
\providecommand{\doarXiv}[1]{\href{https://arxiv.org/abs/#1}{\nolinkurl{https://arxiv.org/abs/#1}}}

\bibitem[{{Argyriou} {et~al.}(2023){Argyriou}, {Glasse}, {Law}, {Labiano}, {{\'A}lvarez-M{\'a}rquez}, {Patapis}, {Kavanagh}, {Gasman}, {Mueller}, {Larson}, {Vandenbussche}, {Glauser}, {Royer}, {Dicken}, {Harkett}, {Sargent}, {Engesser}, {Jones}, {Kendrew}, {Noriega-Crespo}, {Brandl}, {Rieke}, {Wright}, {Lee}, \& {Wells}}]{2023A&A...675A.111A}
{Argyriou}, I., {Glasse}, A., {Law}, D.~R., {et~al.} 2023, \aap, 675, A111, \dodoi{10.1051/0004-6361/202346489}

\bibitem[{{Barklem} {et~al.}(2000){Barklem}, {Piskunov}, \& {O'Mara}}]{BPM}
{Barklem}, P.~S., {Piskunov}, N., \& {O'Mara}, B.~J. 2000, Astron. and Astrophys. Suppl. Ser., 142, 467, \dodoi{10.1051/aas:2000167}

\bibitem[{{Barnes} {et~al.}(2016){Barnes}, {Haswell}, {Staab}, \& {Anglada-Escud{\'e}}}]{2016MNRAS.462.1012B}
{Barnes}, J.~R., {Haswell}, C.~A., {Staab}, D., \& {Anglada-Escud{\'e}}, G. 2016, \mnras, 462, 1012, \dodoi{10.1093/mnras/stw1713}

\bibitem[{Bellotti {et~al.}(2022)Bellotti, Petit, Morin, Hussain, Folsom, Carmona, Delfosse, \& Moutou}]{Bellotti2022Jan}
Bellotti, S., Petit, P., Morin, J., {et~al.} 2022, Astron. Astrophys., 657, A107, \dodoi{10.1051/0004-6361/202141812}

\bibitem[{{Berdyugina} {et~al.}(2003){Berdyugina}, {Solanki}, \& {Frutiger}}]{2003A&A...412..513B}
{Berdyugina}, S.~V., {Solanki}, S.~K., \& {Frutiger}, C. 2003, \aap, 412, 513, \dodoi{10.1051/0004-6361:20031473}

\bibitem[{Bertaux {et~al.}(2014)Bertaux, Lallement, Ferron, Boonne, \& Bodichon}]{Bertaux2014Apr}
Bertaux, J.~L., Lallement, R., Ferron, S., Boonne, C., \& Bodichon, R. 2014, Astron. Astrophys., 564, A46, \dodoi{10.1051/0004-6361/201322383}

\bibitem[{Bixel {et~al.}(2019)Bixel, Rackham, Apai, Espinoza, L{\ifmmode\acute{o}\else\'{o}\fi}pez-Morales, Osip, Jord{\ifmmode\acute{a}\else\'{a}\fi}n, McGruder, \& Weaver}]{Bixel2019Feb}
Bixel, A., Rackham, B.~V., Apai, D., {et~al.} 2019, Astron. J., 157, 68, \dodoi{10.3847/1538-3881/aaf9a3}

\bibitem[{Bouchy {et~al.}(2005)Bouchy, Udry, Mayor, Moutou, Pont, Iribarne, Da~Silva, Ilovaisky, Queloz, Santos, S{\ifmmode\acute{e}\else\'{e}\fi}gransan, \& Zucker}]{Bouchy2005Dec}
Bouchy, F., Udry, S., Mayor, M., {et~al.} 2005, Astron. Astrophys., 444, L15, \dodoi{10.1051/0004-6361:200500201}

\bibitem[{Boyajian {et~al.}(2014)Boyajian, von Braun, Feiden, Huber, Basu, Demarque, Fischer, Schaefer, Mann, White, Maestro, Brewer, Lamell, Spada, López-Morales, Ireland, Farrington, van Belle, Kane, Jones, ten Brummelaar, Ciardi, McAlister, Ridgway, Goldfinger, Turner, \& Sturmann}]{10.1093/mnras/stu2502}
Boyajian, T., von Braun, K., Feiden, G.~A., {et~al.} 2014, Monthly Notices of the Royal Astronomical Society, 447, 846, \dodoi{10.1093/mnras/stu2502}

\bibitem[{Boyajian {et~al.}(2015)Boyajian, von Braun, Feiden, Huber, Basu, Demarque, Fischer, Schaefer, Mann, White, Maestro, Brewer, Lamell, Spada, L{\ifmmode\acute{o}\else\'{o}\fi}pez-Morales, Ireland, Farrington, van Belle, Kane, Jones, Ten~Brummelaar, Ciardi, McAlister, Ridgway, Goldfinger, Turner, \& Sturmann}]{Boyajian2015Feb}
---. 2015, Mon. Not. R. Astron. Soc., 447, 846, \dodoi{10.1093/mnras/stu2502}

\bibitem[{Cauley {et~al.}(2017)Cauley, Redfield, \& Jensen}]{Cauley2017Apr}
Cauley, P.~W., Redfield, S., \& Jensen, A.~G. 2017, Astron. J., 153, 217, \dodoi{10.3847/1538-3881/aa6a15}

\bibitem[{Cegla {et~al.}(2016)Cegla, Lovis, Bourrier, Beeck, Watson, \& Pepe}]{Cegla2016Apr}
Cegla, H.~M., Lovis, C., Bourrier, V., {et~al.} 2016, Astron. Astrophys., 588, A127, \dodoi{10.1051/0004-6361/201527794}

\bibitem[{Cincunegui {et~al.}(2007)Cincunegui, D{\ifmmode\acute{\imath}\else\'{\i}\fi}az, \& Mauas}]{Cincunegui2007Jul}
Cincunegui, C., D{\ifmmode\acute{\imath}\else\'{\i}\fi}az, R.~F., \& Mauas, P. J.~D. 2007, Astron. Astrophys., 469, 309, \dodoi{10.1051/0004-6361:20066503}

\bibitem[{de~Paz {et~al.}(2014)de~Paz, Gallego, Carrasco, Iglesias-P{\ifmmode\acute{a}\else\'{a}\fi}ramo, \& Villar}]{dePaz2014Jun}
de~Paz, A.~G., Gallego, J., Carrasco, E., Iglesias-P{\ifmmode\acute{a}\else\'{a}\fi}ramo, J., \& Villar, V. 2014, Proc. SPIE Int. Soc. Opt. Eng., 9147, \dodoi{10.1117/12.2047825}

\bibitem[{de~Paz {et~al.}(2012)de~Paz, Carrasco, Gallego, S{\'a}nchez, Medina, Garc{\'i}a-Vargas, Arrillaga, Carrera, Castillo-Morales, Castillo-Dom{\'i}nguez, Cedazo, Eliche-Moral, Ferrusca, Gonz{\'a}lez-Guardia, Maldonado, Marino, Mart{\'i}nez-Delgado, Dur{\'a}n, M{\'u}jica, Pascual, P{\'e}rez-Calpena, S{\'a}nchez-Penim, S{\'a}nchez-Blanco, Serena, Tulloch, Villar, Zamorano, y~Nav{\'a}scues, Bertone, Cardiel, Cava, Cenarro, Ch{\'a}vez, Garc{\'i}a, Guichard, G{\'u}zman, Herrero, Hu{\'e}lamo, Hughes, Iglesias, Jim{\'e}nez-Vicente, Aguerri, Mayya, Abreu, Moll{\'a}, Mu{\~n}oz-Tu{\~n}{\'o}n, Peimbert, Peimbert, P{\'e}rez-Gonz{\'a}lez, Montero, Rodr{\'i}guez, Rodr{\'i}guez-Espinosa, Rodr{\'i}guez-Merino, Rosa, S{\'a}nchez-Almeida, Contreras, S{\'a}nchez-Bl{\'a}zquez, S{\'a}nchez, Sarajedini, Silich, Sim{\'o}n, Tenorio-Tagle, Terlevich, Terlevich, Trujillo, Tsamis, \& Vega}]{10.1117/12.925739}
de~Paz, A.~G., Carrasco, E., Gallego, J., {et~al.} 2012, in Ground-based and Airborne Instrumentation for Astronomy IV, ed. I.~S. McLean, S.~K. Ramsay, \& H.~Takami, Vol. 8446, International Society for Optics and Photonics (SPIE), 84464Q, \dodoi{10.1117/12.925739}

\bibitem[{Deming {et~al.}(2025)Deming, Currie, Meadows, \& Peacock}]{article}
Deming, D., Currie, M., Meadows, V., \& Peacock, S. 2025, The Astronomical Journal, 170, 11, \dodoi{10.3847/1538-3881/add531}

\bibitem[{Fournier-Tondreau {et~al.}(2024)Fournier-Tondreau, MacDonald, Radica, Lafreni{\ifmmode\grave{e}\else\`{e}\fi}re, Welbanks, Piaulet, Coulombe, Allart, Morel, Artigau, Albert, Lim, Doyon, Benneke, Rowe, Darveau-Bernier, Cowan, Lewis, Cook, Flagg, Genest, Pelletier, Johnstone, Dang, Kaltenegger, Taylor, \& Turner}]{Fournier-Tondreau2024Feb}
Fournier-Tondreau, M., MacDonald, R.~J., Radica, M., {et~al.} 2024, Mon. Not. R. Astron. Soc., 528, 3354, \dodoi{10.1093/mnras/stad3813}

\bibitem[{Fournier-Tondreau {et~al.}(2025)Fournier-Tondreau, Pan, Morel, Lafrenière, MacDonald, Coulombe, Allart, Albert, Radica, Piaulet-Ghorayeb, Roy, Pelletier, Dang, Doyon, Benneke, Cowan, Darveau-Bernier, Lim, Artigau, Johnstone, Kaltenegger, Taylor, \& Flagg}]{10.1093/mnras/staf489}
Fournier-Tondreau, M., Pan, Y., Morel, K., {et~al.} 2025, Monthly Notices of the Royal Astronomical Society, 539, 422, \dodoi{10.1093/mnras/staf489}

\bibitem[{Fu {et~al.}(2024)Fu, Welbanks, Deming, Inglis, Zhang, Lothringer, Ih, Moses, Schlawin, Knutson, Henry, Greene, Sing, Savel, Kempton, Louie, Line, \& Nixon}]{Fu2024Jul}
Fu, G., Welbanks, L., Deming, D., {et~al.} 2024, Nature, 1, \dodoi{10.1038/s41586-024-07760-y}

\bibitem[{Giles {et~al.}(2017)Giles, Collier~Cameron, \& Haywood}]{Giles2017Dec}
Giles, H. A.~C., Collier~Cameron, A., \& Haywood, R.~D. 2017, Mon. Not. R. Astron. Soc., 472, 1618, \dodoi{10.1093/mnras/stx1931}

\bibitem[{Komori {et~al.}(2025)Komori, Brewer, \& Zhao}]{Komori_2025}
Komori, C., Brewer, J.~M., \& Zhao, L.~L. 2025, The Astronomical Journal, 170, 209, \dodoi{10.3847/1538-3881/adf749}

\bibitem[{{Kurucz}(2014)}]{K14}
{Kurucz}, R.~L. 2014, Robert L. Kurucz on-line database of observed and predicted atomic transitions

\bibitem[{Lafarga {et~al.}(2021)Lafarga, Ribas, Reiners, Quirrenbach, Amado, Caballero, Azzaro, B{\ifmmode\acute{e}\else\'{e}\fi}jar, Cort{\ifmmode\acute{e}\else\'{e}\fi}s-Contreras, Dreizler, Hatzes, Henning, Jeffers, Kaminski, K{\ifmmode\ddot{u}\else\"{u}\fi}rster, Montes, Morales, Oshagh, Rodr{\ifmmode\acute{\imath}\else\'{\i}\fi}guez-L{\ifmmode\acute{o}\else\'{o}\fi}pez, Sch{\ifmmode\ddot{o}\else\"{o}\fi}fer, Schweitzer, \& Zechmeister}]{Lafarga2021Aug}
Lafarga, M., Ribas, I., Reiners, A., {et~al.} 2021, Astron. Astrophys., 652, A28, \dodoi{10.1051/0004-6361/202140605}

\bibitem[{{McCullough} {et~al.}(2014){McCullough}, {Crouzet}, {Deming}, \& {Madhusudhan}}]{2014ApJ...791...55M}
{McCullough}, P.~R., {Crouzet}, N., {Deming}, D., \& {Madhusudhan}, N. 2014, \apj, 791, 55, \dodoi{10.1088/0004-637X/791/1/55}

\bibitem[{{Murphy} {et~al.}(2025){Murphy}, {Beatty}, {Schlawin}, {Bell}, {Radica}, {Kennedy}, {Mehta}, {Welbanks}, {Line}, {Parmentier}, {Greene}, {Mukherjee}, {Fortney}, {Ohno}, {Wiser}, {Arnold}, {Rauscher}, {Edelman}, \& {Rieke}}]{2025AJ....170...61M}
{Murphy}, M.~M., {Beatty}, T.~G., {Schlawin}, E., {et~al.} 2025, \aj, 170, 61, \dodoi{10.3847/1538-3881/addf38}

\bibitem[{{Neff} {et~al.}(1995){Neff}, {O'Neal}, \& {Saar}}]{1995ApJ...452..879N}
{Neff}, J.~E., {O'Neal}, D., \& {Saar}, S.~H. 1995, \apj, 452, 879, \dodoi{10.1086/176356}

\bibitem[{Neilson \& Lester(2013)}]{Neilson2013Jun}
Neilson, H.~R., \& Lester, J.~B. 2013, Astron. Astrophys., 554, A98, \dodoi{10.1051/0004-6361/201321502}

\bibitem[{{O'Brian} {et~al.}(1991){O'Brian}, {Wickliffe}, {Lawler}, {Whaling}, \& {Brault}}]{BWL}
{O'Brian}, T.~R., {Wickliffe}, M.~E., {Lawler}, J.~E., {Whaling}, W., \& {Brault}, J.~W. 1991, Journal of the Optical Society of America B Optical Physics, 8, 1185

\bibitem[{Oshagh {et~al.}(2014)Oshagh, Santos, Ehrenreich, Haghighipour, Figueira, Santerne, \& Montalto}]{Oshagh2014Aug}
Oshagh, M., Santos, N.~C., Ehrenreich, D., {et~al.} 2014, Astron. Astrophys., 568, A99, \dodoi{10.1051/0004-6361/201424059}

\bibitem[{{Pakhomov} {et~al.}(2019){Pakhomov}, {Ryabchikova}, \& {Piskunov}}]{2019ARep...63.1010P}
{Pakhomov}, Y.~V., {Ryabchikova}, T.~A., \& {Piskunov}, N.~E. 2019, Astronomy Reports, 63, 1010, \dodoi{10.1134/S1063772919120047}

\bibitem[{Panwar {et~al.}(2022)Panwar, Désert, Todorov, Bean, Stevenson, Huitson, Fortney, \& Bergmann}]{10.1093/mnras/stac1949}
Panwar, V., Désert, J.-M., Todorov, K.~O., {et~al.} 2022, Monthly Notices of the Royal Astronomical Society, 515, 5018, \dodoi{10.1093/mnras/stac1949}

\bibitem[{Pascual {et~al.}(2024)Pascual, Cardiel, \& Picazo-Sánchez}]{pascual_2024_14186029}
Pascual, S., Cardiel, N., \& Picazo-Sánchez, P. 2024, guaix-ucm/numina: Release v0.35.2, v0.35.2,  Zenodo, \dodoi{10.5281/zenodo.14186029}

\bibitem[{Pietrow {et~al.}(2023)Pietrow, Kiselman, Andriienko, de~la Roche, Baso, \& Calvo}]{Pietrow2023Mar}
Pietrow, A. G.~M., Kiselman, D., Andriienko, O., {et~al.} 2023, Astron. Astrophys., 671, A130, \dodoi{10.1051/0004-6361/202244811}

\bibitem[{{Pont} {et~al.}(2008){Pont}, {Knutson}, {Gilliland}, {Moutou}, \& {Charbonneau}}]{2008MNRAS.385..109P}
{Pont}, F., {Knutson}, H., {Gilliland}, R.~L., {Moutou}, C., \& {Charbonneau}, D. 2008, \mnras, 385, 109, \dodoi{10.1111/j.1365-2966.2008.12852.x}

\bibitem[{{Pont} {et~al.}(2013){Pont}, {Sing}, {Gibson}, {Aigrain}, {Henry}, \& {Husnoo}}]{2013MNRAS.432.2917P}
{Pont}, F., {Sing}, D.~K., {Gibson}, N.~P., {et~al.} 2013, \mnras, 432, 2917, \dodoi{10.1093/mnras/stt651}

\bibitem[{Rackham {et~al.}(2018)Rackham, Apai, \& Giampapa}]{Rackham2018Feb}
Rackham, B.~V., Apai, D., \& Giampapa, M.~S. 2018, Astrophys. J., 853, 122, \dodoi{10.3847/1538-4357/aaa08c}

\bibitem[{Rackham {et~al.}(2023)Rackham, Espinoza, Berdyugina, Korhonen, MacDonald, Montet, Morris, Oshagh, Shapiro, Unruh, Quintana, Zellem, Apai, Barclay, Barstow, Bruno, Carone, Casewell, Cegla, Criscuoli, Fischer, Fournier, Giampapa, Giles, Iyer, Kopp, Kostogryz, Krivova, Mallonn, McGruder, Molaverdikhani, Newton, Panja, Peacock, Reardon, Roettenbacher, Scandariato, Solanki, Stassun, Steiner, Stevenson, Tregloan-Reed, Valio, Wedemeyer, Welbanks, Yu, Alam, Davenport, Deming, Dong, Ducrot, Fisher, Gilbert, Kostov, L{\ifmmode\acute{o}\else\'{o}\fi}pez-Morales, Line, Mo{\ifmmode\check{c}\else\v{c}\fi}nik, Mullally, Paudel, Ribas, \& Valenti}]{Rackham2023Jan}
Rackham, B.~V., Espinoza, N., Berdyugina, S.~V., {et~al.} 2023, RAS Techniques and Instruments, 2, 148, \dodoi{10.1093/rasti/rzad009}

\bibitem[{Rieke {et~al.}(2023)Rieke, Kelly, Misselt, Stansberry, Boyer, Beatty, Egami, Florian, Greene, Hainline, Leisenring, Roellig, Schlawin, Sun, Tinnin, Williams, Willmer, Wilson, Clark, Rohrbach, Brooks, Canipe, Correnti, DiFelice, Gennaro, Girard, Hartig, Hilbert, Koekemoer, Nikolov, Pirzkal, Rest, Robberto, Sunnquist, Telfer, Wu, Ferry, Lewis, Baum, Beichman, Doyon, Dressler, Eisenstein, Ferrarese, Hodapp, Horner, Jaffe, Johnstone, Krist, Martin, McCarthy, Meyer, Rieke, Trauger, \& Young}]{Rieke2023Feb}
Rieke, M.~J., Kelly, D.~M., Misselt, K., {et~al.} 2023, Publ. Astron. Soc. Pac., 135, 028001, \dodoi{10.1088/1538-3873/acac53}

\bibitem[{{Sing} {et~al.}(2009){Sing}, {D{\'e}sert}, {Lecavelier Des Etangs}, {Ballester}, {Vidal-Madjar}, {Parmentier}, {Hebrard}, \& {Henry}}]{2009A&A...505..891S}
{Sing}, D.~K., {D{\'e}sert}, J.~M., {Lecavelier Des Etangs}, A., {et~al.} 2009, \aap, 505, 891, \dodoi{10.1051/0004-6361/200912776}

\bibitem[{{Sing} {et~al.}(2011){Sing}, {Pont}, {Aigrain}, {Charbonneau}, {D{\'e}sert}, {Gibson}, {Gilliland}, {Hayek}, {Henry}, {Knutson}, {Lecavelier Des Etangs}, {Mazeh}, \& {Shporer}}]{2011MNRAS.416.1443S}
{Sing}, D.~K., {Pont}, F., {Aigrain}, S., {et~al.} 2011, \mnras, 416, 1443, \dodoi{10.1111/j.1365-2966.2011.19142.x}

\bibitem[{Solana \& Rubio(2005)}]{Solana2005Aug}
Solana, D.~C., \& Rubio, L. R.~B. 2005, Astron. Astrophys., 439, 687, \dodoi{10.1051/0004-6361:20052720}

\bibitem[{Thompson {et~al.}(2024)Thompson, Biagini, Cracchiolo, Petralia, Changeat, Saba, Morello, Morvan, Micela, \& Tinetti}]{Thompson_2024}
Thompson, A., Biagini, A., Cracchiolo, G., {et~al.} 2024, The Astrophysical Journal, 960, 107, \dodoi{10.3847/1538-4357/ad0369}

\bibitem[{{Wallace} {et~al.}(2005){Wallace}, {Hinkle}, \& {Livingston}}]{2005asus.book.....W}
{Wallace}, L., {Hinkle}, K., \& {Livingston}, W.~C. 2005, {An atlas of sunspot umbral spectra in the visible from 15,000 to 25,500 cm-{\textonesuperior} (3920 to 6664 {\r{A}})}

\bibitem[{{Wilson}(1978)}]{1978ApJ...226..379W}
{Wilson}, O.~C. 1978, \apj, 226, 379, \dodoi{10.1086/156618}

\bibitem[{Wise {et~al.}(2018)Wise, Dodson-Robinson, Bevenour, \& Provini}]{Wise_2018}
Wise, A.~W., Dodson-Robinson, S.~E., Bevenour, K., \& Provini, A. 2018, The Astronomical Journal, 156, 180, \dodoi{10.3847/1538-3881/aadd94}

\bibitem[{W{\ifmmode\ddot{o}\else\"{o}\fi}hl(1971)}]{Wohl1971Feb}
W{\ifmmode\ddot{o}\else\"{o}\fi}hl, H. 1971, Sol. Phys., 16, 362, \dodoi{10.1007/BF00162477}

\bibitem[{Wright {et~al.}(2004)Wright, Marcy, Butler, \& Vogt}]{Wright2004Jun}
Wright, J.~T., Marcy, G.~W., Butler, R.~P., \& Vogt, S.~S. 2004, Astrophys. J. Suppl. Ser., 152, 261, \dodoi{10.1086/386283}

\end{thebibliography}

\end{document}